# Effect of Ag and Zr solutes on dislocation emission from the Σ11(332)[110] symmetric tilt grain boundary in fcc Cu

Valery Borovikov[1], Mikhail I. Mendelev[1] and Alexander H. King[1,2]
[1]*Division of Materials Sciences and Engineering, Ames Laboratory, Ames, IA 50011*
[2]*Department of Materials Science and Engineering, Iowa State University, Ames, IA 50011*

Addition of solutes is commonly used to stabilize nanocrystalline materials against grain growth. However, segregating at grain boundaries, these solutes also affect the process of dislocation nucleation from grain boundaries under applied stress. Using atomistic simulations we demonstrate that the effect of solutes on the dislocation nucleation strongly depends on the distribution of solutes at the grain boundary, which can vary dramatically depending on the solute type. In particular, our results indicate that the solutes with a *smaller* size mismatch can be more effective in suppressing dislocation emission from grain boundaries. Bearing in mind that dislocation slip originating from grain boundaries or their triple junctions is the dominant mechanism of plastic deformation when grain sizes are reduced to the nanoscale, we emphasize the importance of the search for the optimal solute additions, which would stabilize the nanocrystalline material against grain growth and, at the same time, effectively suppress the dislocation nucleation from the grain boundaries.

# 1. Introduction

The grain size of polycrystalline materials strongly affects the plastic deformation behavior [1]. In particular, it is well established that a significant strengthening can be achieved by decreasing grain size [2]. However, nanocrystalline materials are thermally unstable, i.e., a significant grain growth/coarsening is occurring even at relatively low temperatures. In addition, a very small grain size (<15 nm) can result in softening due to the crossover from the dislocation-mediated plasticity to the grain boundary (GB) mediated plasticity (grain boundary sliding/rotation) [3,4]. The addition of immiscible solutes, which tend to segregate at grain boundaries, can stabilize nanocrystalline material against the grain growth [5-7] and also suppress grain boundary mediated plasticity mechanisms [8]. Keeping in mind that dislocation emission from grain boundaries is a key deformation mechanism for a wide range of grain sizes [9,10], it would be ideal, if the solutes could also suppress the nucleation of the dislocations from grain boundaries [11].

Recently we demonstrated that solute atoms can have a very strong effect on the process of dislocation nucleation from grain boundaries [11]. In particular we observed a significant increase in the yield stress under applied tensile loading due to the effects of both oversized (Ag in Cu) or undersized (Cu in Ag) solutes. On the other hand, it was also argued that the substitutional solutes with a larger size mismatch the most effectively increase the yield strength of nanocrystalline Cu [12,13]. In particular, it was shown that the yield stress gradually increases with increased the size mismatch [13]. Comparison of the data obtained in [11] and [12,13] leads to the conclusion that the atomic mismatch is not the only decisive factor governing the effect of the solutes atoms. Therefore, further investigation is needed. In the present manuscript, we report the results of the molecular dynamics (MD) simulation of the effects of Ag and Zr solutes on the dislocation nucleation from the $\Sigma 11(332)[110]$ symmetric tilt grain boundary (STGB) in Cu. We demonstrate that the effect of solutes strongly depends on their segregation pattern and solute atoms with a *smaller* size mismatch (Ag in Cu) can suppress the nucleation of dislocations from the grain boundary (increasing the yield stress) more effectively than solute atoms with a larger mismatch (Zr in Cu).

The rest of the paper is organized as follows. First, we will describe the simulation geometry/methods and the employed semi-empirical potential of the interatomic interaction. Next, we will discuss the differences in the segregation patterns of the Ag and Zr solutes in the $\Sigma 11(332)[110]$ symmetric tilt grain boundary STGB. Finally, we will show how these differences affect the dependence of the yield stress on the solute concentration.

# 2. Simulation geometry and interatomic interaction

The effect of two solutes in Cu were considered: Ag and Zr. Both solutes have larger atomic radius (144 pm for Ag and 160 pm for Zr) than does Cu (128 pm). Since MD simulation of plastic deformation requires utilizing large simulation cells containing at least tens of thousands atoms during rather long simulation time, employing of semi-empirical potentials is the only reasonable option. Unfortunately, currently there is no reliable semi-empirical potential for the Cu-Ag-Zr ternary system. Instead, there are well tested potentials for the binary alloys. For example, an embedded atom method (EAM) potential [14] for the Cu-Ag alloys was developed in [15] and a Finnis-Sinclair (FS) potential [16] for the Cu-Zr alloys was developed in [17]. However, the Cu potential from [14] leads to 2.8 meV/Å$^2$ and 10.1 meV/Å$^2$ for the stable and unstable stacking fault energies, respectively, and the Cu potential from [17] leads to 2.4 meV/Å$^2$ and 16.6 meV/Å$^2$. Both stable and unstable stacking fault energies affect the dislocation nucleation from GBs [18]. Obviously, to compare the effects of Ag and Zr additions on the dislocation nucleation in Cu we need the potentials with similar Cu properties. This precondition will be met if the Cu potential in the Cu-Zr potential from [14] is replaced by the MCu31 potential developed in [18]. This potential leads to 2.8 meV/Å$^2$ and 10.1 meV/Å$^2$ for the stable and unstable stacking fault energies, respectively; just like the Cu-Ag potential does.

To further check if the chosen Cu potentials lead to similar behavior during the plastic deformation in the pure Cu we performed MD simulation at T=300 K utilizing a simple bi-crystal geometry in which the emission of dislocations from a grain boundary is the only active mechanism of plastic deformation. The simulation cell has two Σ11(332)[110] symmetric tilt grain boundaries which contain the E structural units [19]. This choice of the studied GB is related to the observation that a subset of symmetric tilt grain boundaries within the <110> family, which contain the E structural units, are able to emit dislocations under lower applied stresses, compared to other grain boundaries [20,21]. All simulations were carried out using the LAMMPS simulation package [22] and the visualization of the simulation snapshots was performed using the OVITO package [23]. The simulation cell is shown in Fig. 1. Periodic boundary conditions were used in all three directions. The simulation cell size was sufficiently large in all three directions (greater than 16 nm in the directions parallel to GB plane, and greater than 32 nm in the direction normal to GB plane) in order to minimize the effect of periodic boundary conditions on dislocation nucleation [20,24]. Uniaxial tensile loading simulations were carried out with a constant engineering strain rate of $10^8$ s$^{-1}$ applied in the z direction (normal to the grain boundary plane), while the stresses in the other two directions were held at zero. Other details of the simulation were identical to those used in [11].

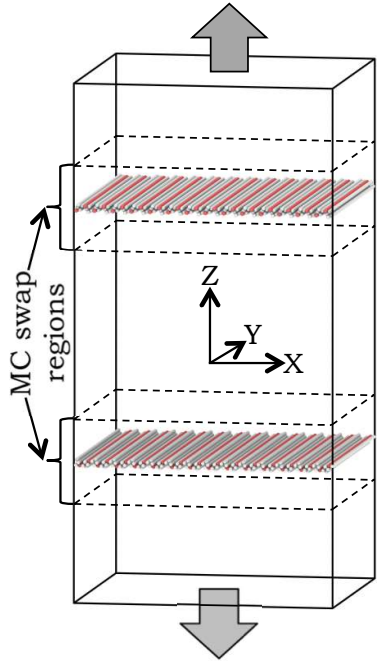

Figure 1. The bi-crystal simulation system; the MC swap regions are indicated by dashed lines. The arrows indicate the direction of the applied deformation.

At the onset of the deformation the stress gradually increases with increasing applied strain but eventually it rather abruptly drops (see Fig. 2), when the first dislocation is emitted. This peak stress value was considered as the effective yield stress in the present study. Figure 2 shows the data obtained for 3 Cu potentials discussed above. The Cu potential taken from the Cu-Ag potential and MCu31 indeed lead to similar curves. On the other hand, the Cu potential taken from the Cu-Zr potential leads to much higher value of the yield stress which is associated with the larger value of the unstable staking fault energy (see the discussion in [11]).

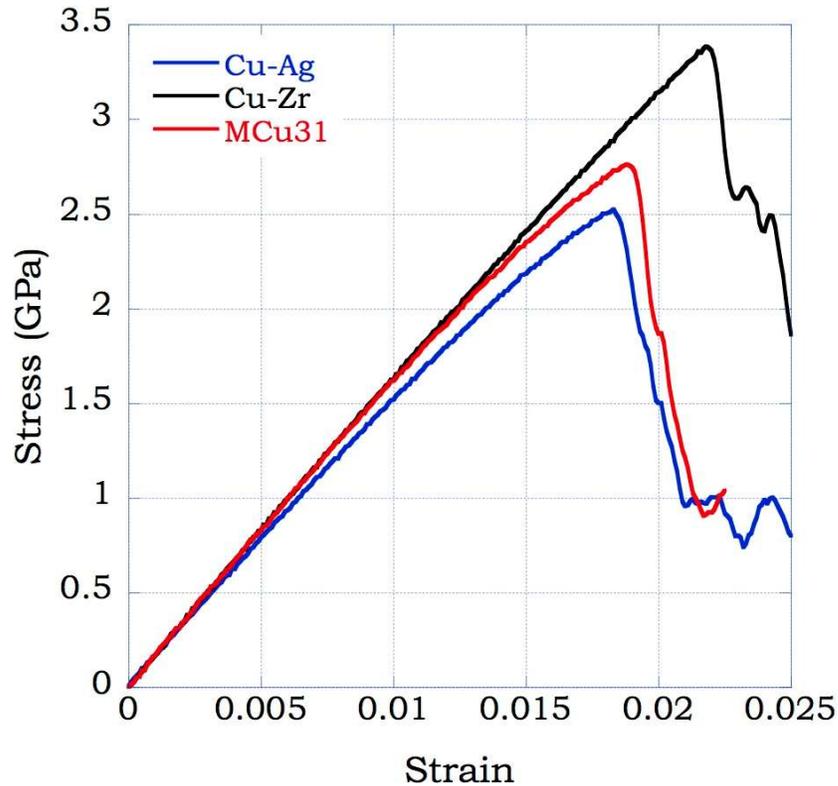

Figure 2. Stress-strain curves for the Cu potentials from [15,17,18].

Since the Cu-Zr FS potential from [17] has been intensively tested in MD simulation we further only refitted the Cu-Zr cross pair potential. The details of the potential development procedure can be found in [25]. All results for the Cu-Zr alloy reported in the rest of the present manuscript were obtained using the new Cu-Zr potential. This potential can be found in [25,26].

## 3. Solute segregation at the grain boundary

Both Ag and Zr atoms are much larger than the Cu atoms; therefore, they both introduce stress fields when they substitute Cu matrix atoms. However, these fields are rather different even in the bulk. As can be seen in Fig. 3a, the single Ag solute is under strong compression and its Cu nearest neighbors are under much smaller compression. On contrary, Fig. 3b demonstrates that the stress distribution around the Zr solute is more complicated: the solute itself is under tension, while the shell of Cu nearest-neighbors is under pretty large compression. We also note that the Ag and Zr solutes have very different dipole interaction energies in the bulk. The Zr solutes are characterized by positive dipole interaction energy, $E_d = 0.29$ eV, which corresponds to repulsion. On the other hand, the Ag solutes have negative dipole interaction energy, $E_d = -0.09$ eV, which corresponds to attraction.

At small concentrations, both Ag and Zr solutes segregate exclusively at the A' sites of the E structural units in the Σ11(332)[110] STGB (see Figs. 5c and 5d) because these sites are under tension as was shown in [27]. The substitution of the Cu atom by the Ag atom at the A' site changes the stress sign at this site: it is now under compression. The substitution of the Cu atom by the Zr atom at the A' leads to a more complex effect: the stress at the site becomes very small but the compressive stress of several atoms on the nearby plane becomes larger.

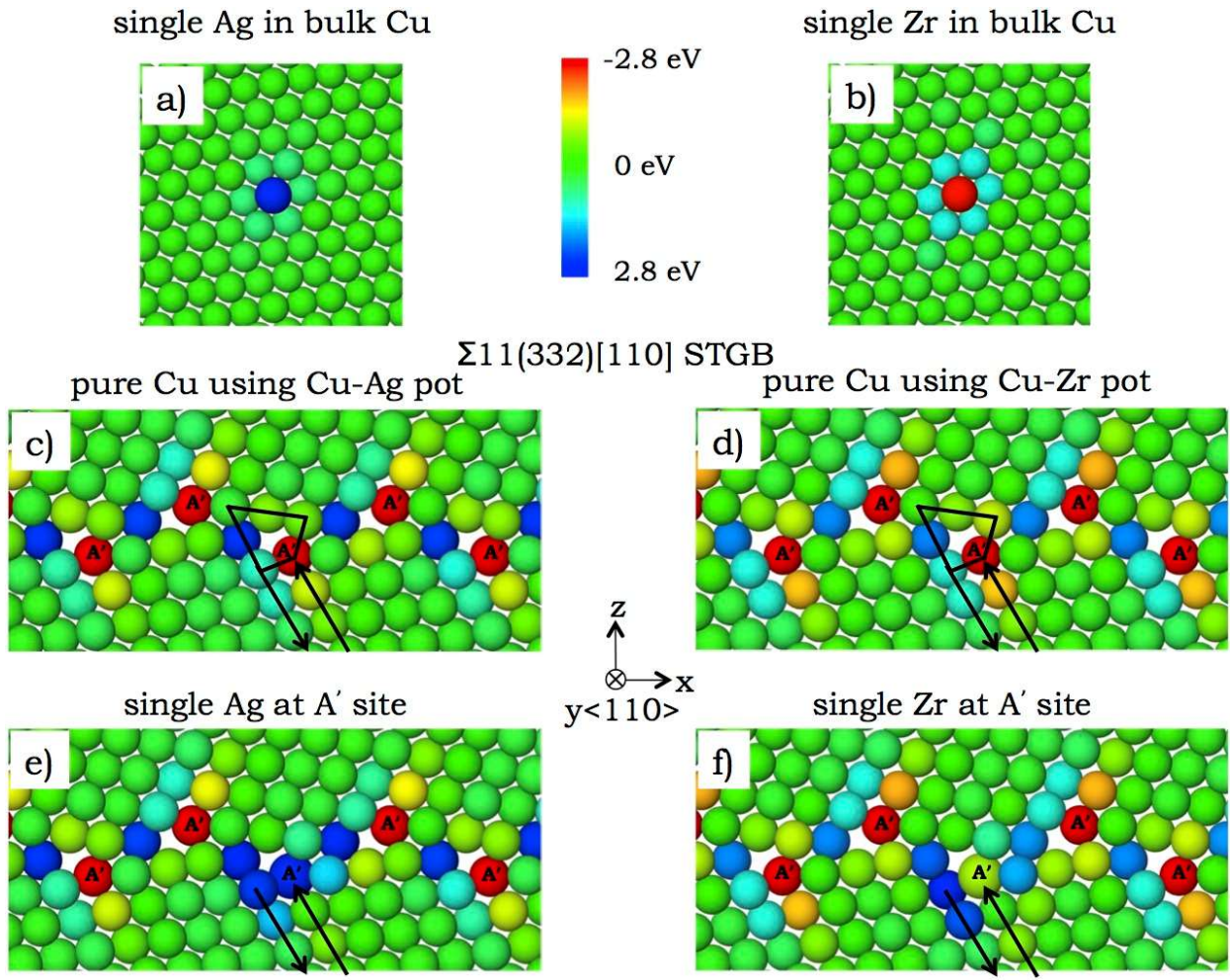

Figure 3. Distribution of stresses around: a) single Ag solute in bulk Cu; b) single Zr solute in bulk Cu; c) Σ11(332)[110] STGB, using the Cu-Ag potential; d) Σ11(332)[110] STGB, using the Cu-Zr potential; e) Σ11(332)[110] STGB with a single Ag solute at the A' site; f) Σ11(332)[110] STGB with a single Zr solute at the A' site. The atoms are colored according to the trace of the atomic stresses. The arrows schematically show the direction of displacements of two atomic planes

during leading partial dislocation emission from the E unit. The black polygons in c) and d) schematically indicate the E units.

The differences in segregation behavior of Ag and Zr solutes become even more pronounced at larger solute concentration. In order to simulate their segregation the solute atoms were initially introduced into the simulation cell by replacing a number of matrix (Cu) atoms in the swap regions near the grain boundaries (see Fig. 1). Next, the hybrid Monte Carlo/ molecular dynamics (MC/MD) simulation was carried out at T=300 K and zero applied stress, in order to obtain an equilibrium solute distribution [28]. The simulation technique was identical to that used in [11]. Figure 4 shows a few representative examples of the solute distributions we observed. At low solute concentrations (< 0.3%), both the Ag and Zr solutes segregate almost exclusively at the A′ sites. However, because the Zr solutes repeal from each other, the segregation at the A′ sites stops at the concentration ~0.28%, when *every other* A′ site along the [110] tilt axis is filled. At higher solute concentrations, the Zr solutes segregate at other sites further away from the grain boundary. On contrary, in the case of Ag solutes, segregation at the A′ sites continues with increasing solute concentration until *every* A′ site along the [110] tilt axis is filled (the corresponding Ag solute concentration is ~0.56%). Then the solutes begin segregating at other sites. As a result, the distribution of the Ag solutes at the GB is more compact at high solute concentrations, compared to the corresponding distribution of the Zr solutes (see Fig. 4).

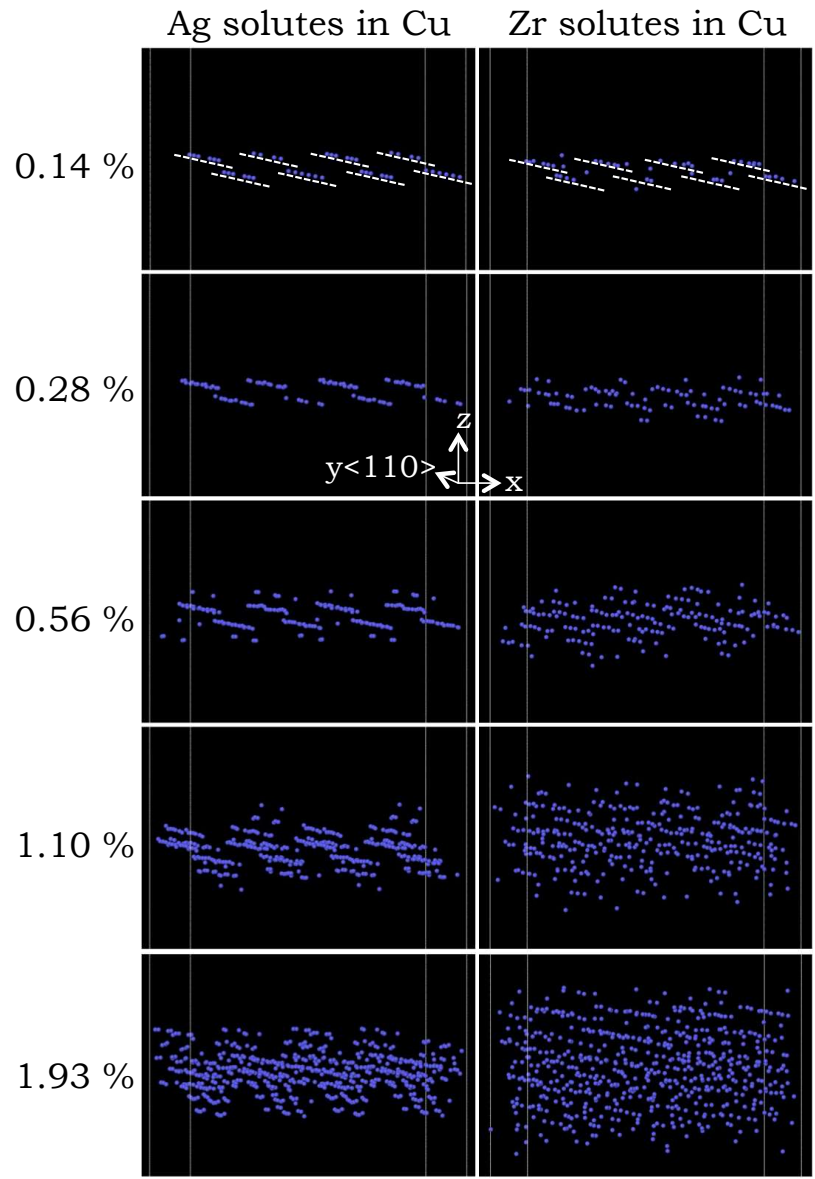

Figure 4. Representative examples of the solute distributions at Σ11(332)[110] STGB in Cu for the cases of Ag and Zr solutes. Only solute atoms are shown. Note that all Ag atoms in the left top image are segregated in the A' sites.

Figure 5 shows how the addition of solutes modifies the GB structure. Clearly the effect of the Ag solutes is much smaller than the effect of the Zr solutes. While segregating rather compactly, the Ag solutes do not considerably change the GB structure. At small concentrations, the Zr solutes also do not change the GB structure but at concentrations larger than ~1.1 %, they make it much more disordered in spite of the fact that the Zr solutes are wider distributed in the direction normal to the GB comparing to the Ag solutes. The GB structure at 1.93 % looks like amorphous with some remains of the initial crystalline structure (which we will discuss in the next section). This effect has been observed in both experiment and MD simulation in [29,30] (although the the original Cu-Zr potential from [17] was employed in [30]).

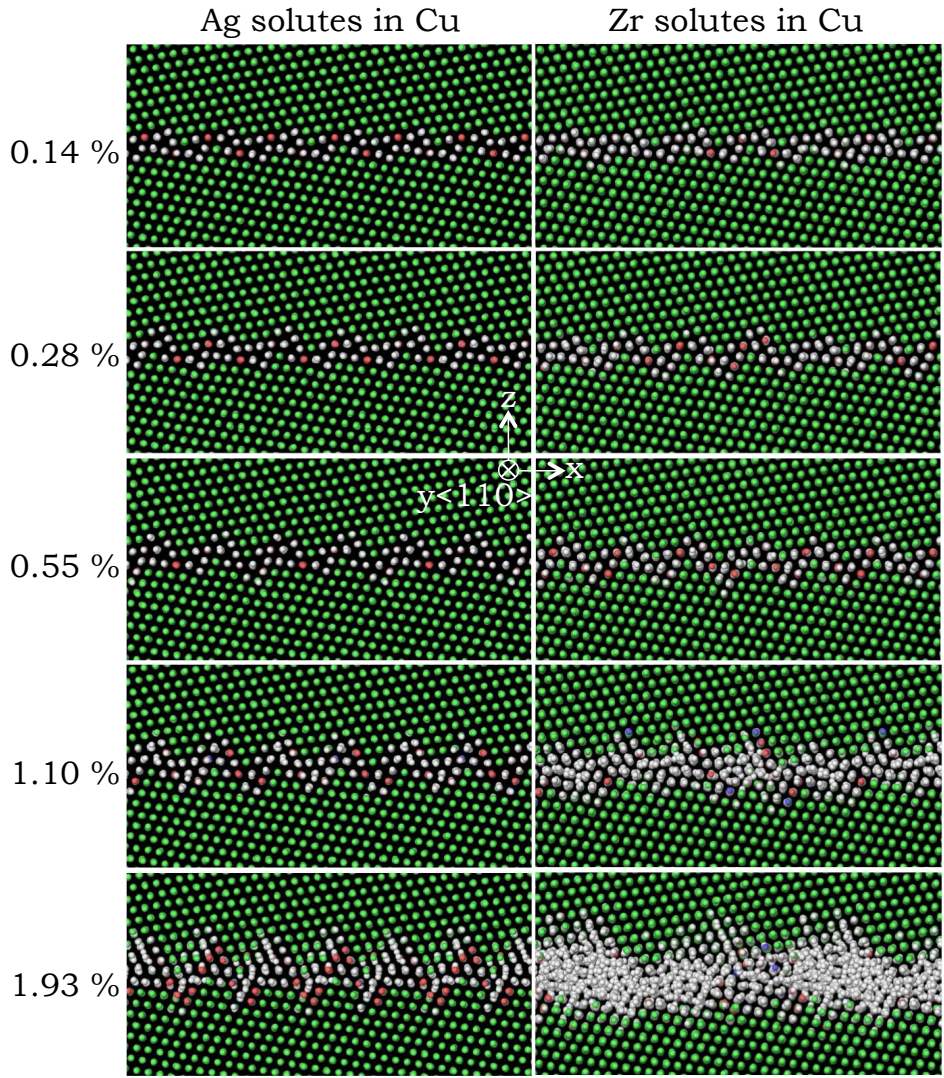

Figure 5. Disorder in the Σ11(332)[110] STGB at different solute contents. The atoms are colored according to the common neighbor analysis (CNA) [31,32]. The color-coding is as follows: green – FCC, red – HCP, grey – others. The size of both solutes and solvents is decreased and made the same to highlight the increase in disorder with increasing solute concentration.

## 4. Effect of solutes on the dislocation nucleation from the GB

The MD simulation of the uniaxial deformation of the simulation cells containing solutes was identical to that for pure Cu (see Section 2). Figure 6 shows the dependence of the yield stress on the solute concentration for both solutes we considered in this study. At low solute concentrations, addition of both Ag and Zr solutes leads to increase in the yield stress. In spite of the fact that the atomic mismatch is larger in the case of the Zr solutes, the effect of Ag seems to be slightly larger. At higher concentrations, the yield stress increases with increasing Ag concentration although it considerably fluctuates after the Ag atoms substitute approximately 30% of the Cu atoms at the grain boundary (which corresponds to ~0.83% total concentration). In the case of the Zr solutes, the yield stress reaches a peak when the Zr atoms substitute approximately 20% of the Cu atoms at the grain boundary (which corresponds to ~0.55% total concentration) and then gradually decreases such that the strengthening effect is completely eliminated at higher Zr concentrations. The observed phenomena are in vivid contradiction with the intuitive assumption that the effect of solutes on the tensile strength will increase with the increasing atomic mismatch.

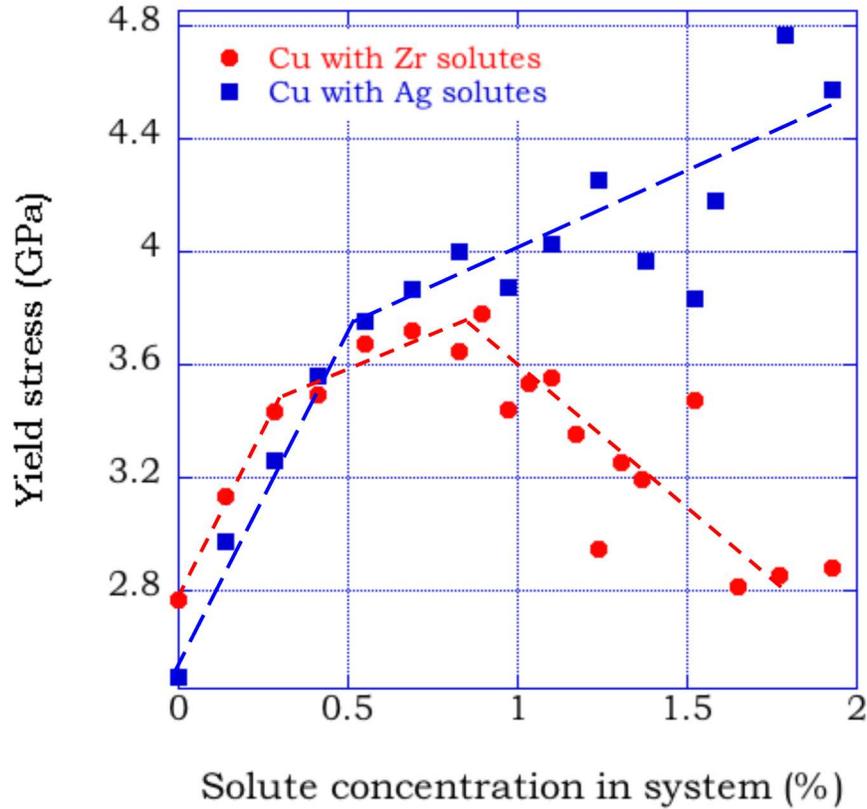

Figure 6. The dependence of the yield stress on the solute concentration. The dashed lines schematically indicate the trends observed at different solute concentrations.

To explain this unexpected result, we need to understand how both solutes affect the mechanism of dislocation nucleation from the GB under investigation. We start from small concentrations when both Ag and Zr solutes segregate exclusively at the A′ sites of the E structural units (see Fig. 4). Recall that the first dislocations are emitted from the E units under applied tensile loading [33]. In the course of the dislocation emission, the plane of atoms which includes the A′ site shifts towards the E unit, while the adjacent plane which includes the atoms under compression, shifts away from the E unit (see Figs. 3c-d). The effect of the Ag solutes on the dislocation emission is pretty straightforward: they change the stress at the A′ sites from tensile to compressive which makes more difficult for the plane of atoms which includes this site to move towards the E unit. This excludes this site from the possible places where a dislocation can nucleate. The effect of the Zr solutes is similar: they also change the stress at the A′ sites in a similar fashion. Contrary to the Ag solutes they do not create a large compressive stress there but yet they also exclude the sites where they segregate from the possible places of the dislocation nucleation.

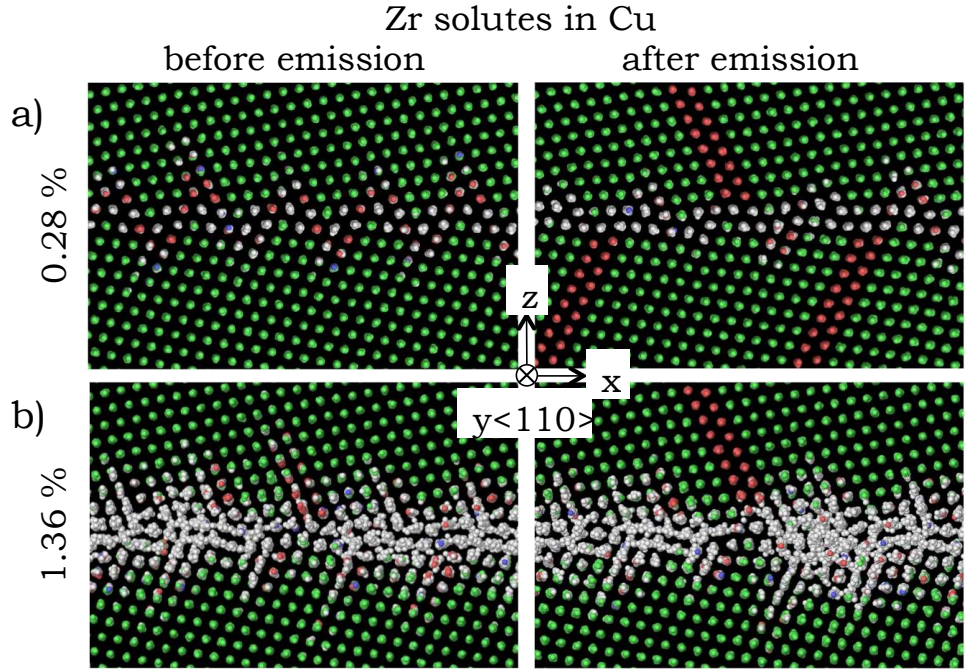

Figure 7. The change of the Σ11(332)[110] STGB structure associated with the dislocation emission. The atoms are colored according to the common neighbor analysis (CNA) [31,32]. The color-coding is as follows: green – FCC, red – HCP, grey – others. The size of solutes and solvents is decreased and made the same to highlight the changes in the GB structure associated with the dislocation emission.

At larger Ag concentrations the solutes keep segregating at the A′ sites, which provides further increase in the yield stress. However, at ~0.56 %, all A′ sites are occupied and the Ag solutes begin segregating at other sites (further away from the GB), which have a weaker effect on dislocation emission. This explains the change in the slope of the yield stress vs. Ag concentration in Fig. 6. Moreover, there are two segregation sites which have a similar segregation energies (a more detailed discussion can be found in [11]) and solute interaction may lead to a complex formation such that it is not obvious that the MC/MD simulation at concentrations larger than 0.56 % always leads to the most energetically favorable solute distribution. This explains the noisy character of the yield stress vs. Ag concentration shown in Fig. 6.

In the case of Zr, the segregation at the A′ sites stops at the concentration ~0.28%, when *every other* A′ site along the [110] tilt axis is filled with solutes. At higher Zr concentrations, the solutes begin segregating at the sites located further away from the GB, which have less influence on the dislocation nucleation. This explains why the yield stress grows slower (with increasing solute concentration) for the solute concentrations between ~0.28 % and ~0.8 %. Finally the decrease in the yield stress for the Zr concentration larger than ~0.8 % can be explained as follows. When the dislocation is emitted from the E unit surrounded by the periodic GB region the displacements of the atoms located near by the source cannot be large, which makes it difficult to emit the dislocation. This is always the case at low solute concentrations (see Fig. 7a). At larger Zr concentrations, the amorphization of the GB starts. Under the applied stress the dislocations still nucleate in the E units, but now these E units are surrounded by the amorphous regions (see

Fig. 7b) where it is easier for the atoms to displace in order to accommodate the atomic shifts associated with the dislocation emission.

## 5. Conclusions

In the present study we performed the MD simulation of the effect of the Ag and Zr solutes on the dislocation nucleation from the $\Sigma 11(332)[110]$ STGB in Cu. In spite the fact that the Ag solutes have smaller atomic radii, compared with the Zr solutes, the effect of the Ag solute addition on the tensile strength is larger than the corresponding effect of the Zr solutes, especially at large solute concentrations. This was explained by the fact that the Zr solutes cannot segregate at the adjacent A' sites, and, therefore, at relatively low solute concentration they will begin segregating at the sites which have a smaller effect on the dislocation nucleation. Moreover, the further addition of Zr leads to the amorphization of the GB and reduces the yield stress.

While a number of recent studies demonstrated that Zr is an excellent candidate for stabilizing nanocrystalline Cu against the grain growth at elevated temperatures [29,34,35], our results indicate that it may not be very effective in suppressing the dislocation emission from grain boundaries under applied tensile loading (compared, for example, to Ag). Bearing in mind that dislocation slip originating from grain boundaries or their triple junctions is the dominant mechanism of plastic deformation when grain sizes are reduced to the nanoscale, we emphasize the importance of the search for the optimal solute additions, which would stabilize the nanocrystalline material against grain growth and, at the same time, effectively suppress the dislocation nucleation from the grain boundaries.

**Acknowledgements:**

This work was supported by the U.S. Department of Energy, Office of Science, Basic Energy Sciences, Materials Science and Engineering Division. The research was performed at Ames Laboratory, which is operated for the U.S. DOE by Iowa State University under contract # DE-AC02-07CH11358

**References:**

[1]     E. N. Hahn and M. A. Meyers, Materials Science and Engineering a-Structural Materials Properties Microstructure and Processing **646**, 101 (2015).
[2]     R. W. Armstrong, Materials Transactions **55**, 2 (2014).
[3]     J. Schiotz and K. W. Jacobsen, Science **301**, 1357 (2003).
[4]     A. S. Argon and S. Yip, Philosophical Magazine Letters **86**, 713 (2006).
[5]     T. Chookajorn, H. A. Murdoch, and C. A. Schuh, Science **337**, 951 (2012).
[6]     T. Frolov, K. A. Darling, L. J. Kecskes, and Y. Mishin, Acta Materialia **60**, 2158 (2012).
[7]     K. A. Darling, M. A. Tschopp, B. K. VanLeeuwen, M. A. Atwater, and Z. K. Liu, Computational Materials Science **84**, 255 (2014).
[8]     P. C. Millett, R. P. Selvam, and A. Saxena, Materials Science and Engineering A-Structural Materials Properties Microstructure and Processing **431**, 92 (2006).
[9]     B. Chen *et al.*, Science **338**, 1448 (2012).
[10]    J. Chen, L. Lu, and K. Lu, Scripta Materialia **54**, 1913 (2006).

[11] V. Borovikov, M. I. Mendelev, and A. H. King, International Journal of Plasticity **90**, 146 (2017).
[12] S. Ozerinc, K. P. Tai, N. Q. Vo, P. Bellon, R. S. Averback, and W. P. King, Scripta Materialia **67**, 720 (2012).
[13] N. Q. Vo, J. Schafer, R. S. Averback, K. Albe, Y. Ashkenazy, and P. Bellon, Scripta Materialia **65**, 660 (2011).
[14] M. S. Daw and M. I. Baskes, Physical Review B **29**, 6443 (1984).
[15] P. L. Williams, Y. Mishin, and J. C. Hamilton, Modelling and Simulation in Materials Science and Engineering **14**, 817 (2006).
[16] M. W. Finnis and J. E. Sinclair, Philosophical Magazine a-Physics of Condensed Matter Structure Defects and Mechanical Properties **50**, 45 (1984).
[17] M. I. Mendelev, M. J. Kramer, R. T. Ott, D. J. Sordelet, D. Yagodin, and P. Popel, Philos. Mag. **89**, 967 (2009).
[18] V. Borovikov, M. I. Mendelev, and A. H. King, Modelling and Simulation in Materials Science and Engineering **24**, 14, 085017 (2016).
[19] J. D. Rittner and D. N. Seidman, Physical Review B **54**, 6999 (1996).
[20] D. E. Spearot, M. A. Tschopp, K. I. Jacob, and D. L. McDowell, Acta Materialia **55**, 705 (2007).
[21] M. A. Tschopp, G. J. Tucker, and D. L. McDowell, Acta Materialia **55**, 3959 (2007).
[22] S. Plimpton, J. Comput. Phys. **117**, 1 (1995).
[23] A. Stukowski, Modelling and Simulation in Materials Science and Engineering **18**, 015012 (2010).
[24] D. E. Spearot, K. I. Jacob, and D. L. McDowell, International Journal of Plasticity **23**, 143 (2007).
[25] Supplemental Material.
[26] Interatomic Potentials Repository Project, http://www.ctcms.nist.gov/potentials.
[27] O. H. Duparc, A. Larere, B. Lezzar, O. Khalfallah, and V. Paidar, J. Mater. Sci. **40**, 3169 (2005).
[28] B. Sadigh, P. Erhart, A. Stukowski, A. Caro, E. Martinez, and L. Zepeda-Ruiz, Physical Review B **85**, 184203 (2012).
[29] A. Khalajhedayati and T. J. Rupert, Jom **67**, 2788 (2015).
[30] Z. L. Pan and T. J. Rupert, Physical Review B **93**, 15, 134113 (2016).
[31] J. D. Honeycutt and H. C. Andersen, Journal of Physical Chemistry **91**, 4950 (1987).
[32] A. Stukowski, Modelling and Simulation in Materials Science and Engineering **20**, 045021 (2012).
[33] D. E. Spearot, Mech. Res. Commun. **35**, 81 (2008).
[34] M. A. Atwater, R. O. Scattergood, and C. C. Koch, Materials Science and Engineering a-Structural Materials Properties Microstructure and Processing **559**, 250 (2013).
[35] T. J. Rupert, Current Opinion in Solid State & Materials Science **20**, 257 (2016).